# ThAME: 3D Memory-Enabled Heterogeneous Accelerator for LLM Mixture of Experts

Pratyush Dhingra, *Graduate Student Member, IEEE,* Pramit Kumar Pal, *Graduate Student Member, IEEE,* Janardhan Rao Doppa, *Senior Member, IEEE, and* Partha Pratim Pande, *Fellow, IEEE*

***Abstract*—Mixture of Experts (MoE) architectures have emerged as a dominant paradigm for scaling Large Language Models (LLMs). However, MoE inference on conventional hardware is constrained by three fundamental bottlenecks. These encompass the massive memory bandwidth required to fetch non-contiguous expert weights, the non-deterministic scatter-gather traffic generated by input-dependent token routing, and the tail-latency dependency imposed by synchronous expert output aggregation. To address these challenges, we propose ThAME, a three-dimensional (3D) heterogeneous multi-chiplet architecture for MoE inference. ThAME employs Ferroelectric Field-Effect Transistor (FeFET)-based non-volatile and DRAM-based volatile memory chiplets with a co-designed compute mapping strategy that aligns the distinct computational profiles of attention mechanisms and expert routing. Furthermore, we design a specialized Network-on-Chip communication backbone optimized to mitigate the bottlenecks associated with non-deterministic token routing traffic across the combinatorial space of input-dependent MoE traffic patterns. Experimental results demonstrate that ThAME outperforms state-of-the-art counterparts by up to 15.7× in terms of speedup and improves energy efficiency by up to 9.8×.**

## I. INTRODUCTION

Large Language Models (LLMs) have become a foundational paradigm in modern AI, delivering state-of-the-art performance across diverse applications including language, code, vision, and robotics [1]. Built on the transformer architecture, they leverage self-attention mechanisms to capture long-range dependencies and achieve exceptional accuracy and generalization capabilities [1]. Recently, the Mixture of Experts (MoE) architecture is rapidly becoming the preferred practical paradigm to efficiently scale these models to billions of parameters [2]. By activating a small subset of specialized experts per token, MoE-based LLMs achieve high representational power of dense feed-forward networks through sparse computation [2].

However, this sparse computation in MoE-based LLM inference poses significant challenges. First, MoE models demand enormous memory capacity to store the vast number of static expert weights. During the autoregressive decode phase, each generated token necessitates fetching large, non-contiguous expert weight matrices from memory. This memory-bound nature saturates the available bandwidth on conventional computing platforms such as GPUs, leaving compute units underutilized [3], [4]. Second, the input-dependent nature of MoE execution produces irregular dataflow. Each token is dynamically routed to a subset of relevant experts that are spatially distributed across computing cores, generating non-deterministic, bursty scatter-gather communication patterns that change with every input batch [2]. Third, MoE inference requires synchronous aggregation of outputs from all selected experts, imposing a strict tail-latency dependency where a single congested expert can stall the throughput of the entire batch.

To address the memory wall, memory-centric architectures, particularly Processing-in-Memory (PIM) and Processing-Near-Memory (PNM), have emerged as a promising paradigm [5] These architectures perform energy-efficient computation within or near the memory modules to drastically reduce the data movement overhead. However, MoE workloads exhibit significant heterogeneity in their computational kernels. The attention mechanism requires read-modify-write operations, demanding high endurance [6], [7], [8], [9]. In contrast, the expert layers involve multiplications with massive, static pre-

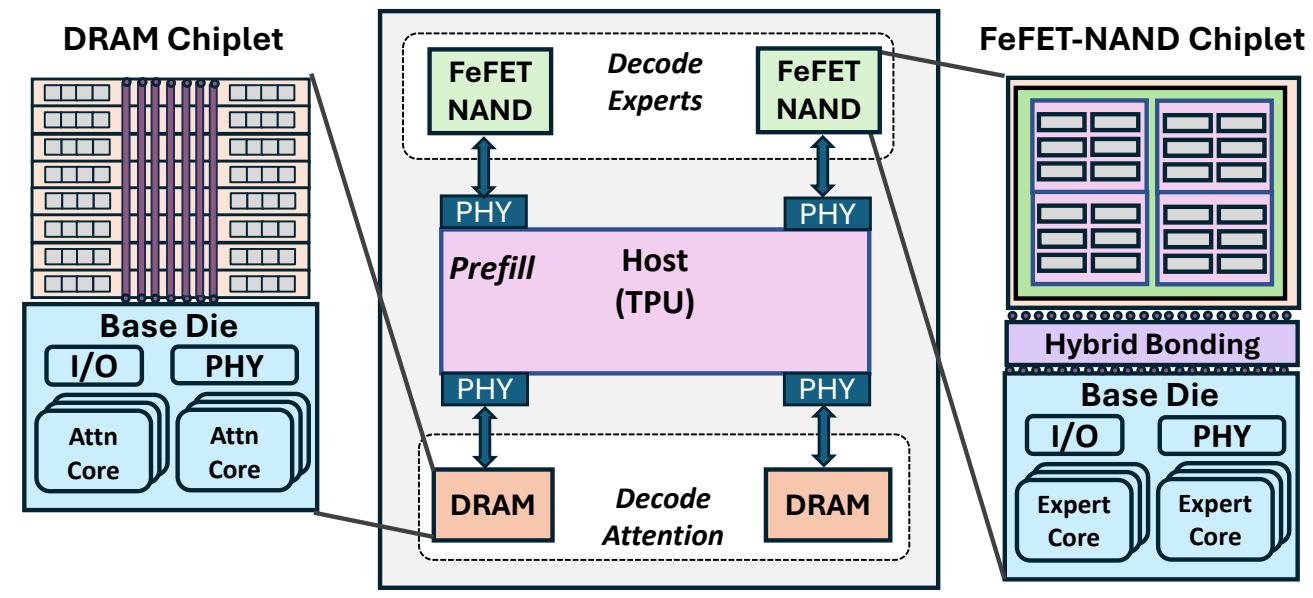

**Fig. 1:** High-level architecture of ThAME. The chiplets are connected through the interposer via UCIe. This figure is for illustration purposes only.

This work was supported by the US National Science Foundation (NSF) grant CSR-2308530, the Army Research Office under Grant ARO-W911NF-24-1-0240 and the U.S. Department of Energy, Office of Science, Advanced Scientific Computing Research program under project 84245—Democratization of Co-design for Energy-Efficient Heterogeneous Computing (DeCoDe) at Pacific Northwest National Laboratory (PNNL). PNNL is a multi-program national laboratory operated for the U.S. Department of Energy (DOE) by Battelle Memorial Institute under Contract No. DE-AC05-76RL01830. *(Corresponding author: Pratyush Dhingra.)*

The authors are with the Department of Electrical Engineering and Computer Science, Washington State University, Pullman, WA 99163 USA (e-mail: pratyush.dhingra@wsu.edu; pramit.pal@wsu.edu; jana.doppa@wsu.edu; pande@wsu.edu).

trained weights, requiring high storage density [6], [7], [8], [9]. A single memory technology cannot efficiently address the demands imposed by MoE workloads due to trade-offs in terms of power, area, latency, endurance, and device variability [10]. Consequently, a heterogeneous memory substrate along with a co-designed solution to map computational kernels on hardware platform is necessary to accelerate MoE inference [8], [9], [11].

In this paper, we propose a novel heterogeneous multi-chiplet architecture, referred to as ThAME. Fig. 1 shows the ThAME architecture that consists of three-dimensional (3D) PNM chiplets along with the host chiplet in a heterogeneous 2.5D system. ThAME leverages 3D vertical stacking of memory layers to overcome the storage requirements of MoE models. Specifically, we utilize 3D DRAM-based volatile memory chiplets for the attention computation and Ferroelectric Field-Effect Transistor (FeFET)-based NVM chiplets to provide the high storage density required for the expert layers. The CMOS base die within the memory chiplets incorporate computing cores to enable PNM and address the extreme memory bandwidth demands imposed by MoE models.

However, optimizing the compute and memory substrate addresses only part of the problem. A significant remaining challenge is the non-deterministic communication patterns which are specific to each batch of inputs. During MoE inference, experts are distributed across the computing cores within the base die, necessitating dynamic communication among the constituent cores and generating unpredictable, bursty traffic [12], [13]. Existing accelerators adopt standard interconnects designed for static dataflow patterns that fail to address the network contention arising from non-deterministic MoE token routing [8], [14], [15]. To address this challenge, we develop a specialized intra-chiplet network-on-chip (NoC) that interconnects the compute cores in the memory-chiplet base die. This NoC is designed to mitigate the bottlenecks associated with non-deterministic token routing traffic across the combinatorial space of input-dependent MoE traffic patterns. The key contributions of this paper include:

- We propose ThAME, a heterogeneous architecture that pairs a DRAM and FeFET memory substrate with a co-designed mapping strategy, delivering the high-density storage and bandwidth demanded by MoE workloads.
- We develop a novel hierarchical NoC communication backbone optimized for the non-deterministic traffic patterns, bursty congestion, and token-routing bottlenecks associated with MoE inference over a batch of inputs.
- We demonstrate that ThAME balances both compute and communication to optimize system utilization, achieving up to 15.7× speedup and improving energy efficiency by 9.8× compared to state-of-the-art hardware accelerators.

The rest of this paper is organized as follows: Section II discusses prior work and provides relevant background. Section III outlines the problem statement, and Section IV elaborates on the proposed architecture and innovations. Section V presents the experimental results, and Section VI concludes the paper.

## II. Background & Related Work

In this section, we discuss MoE LLM inference and provide relevant background on existing memory-centric heterogeneous accelerators designed for LLMs and MoE.

### *A. LLM Inference with Transformers*

Inference in transformer based LLMs operates in two distinct phases: *prefill* and *decode* [1], [4]. The prefill phase processes the input prompt in parallel to populate the Key ($K$) and Value ($V$) tensors for all tokens and stores them in the Key-Value ($KV$) cache. In contrast, the model generates output tokens one at a time in an autoregressive loop in the decode phase [16]. While the prefill phase is typically compute-bound, the decode phase is memory-bandwidth bound given the low arithmetic intensity (defined as the ratio of computations to memory access) [16]. Consequently, for every generated token during the decode phase, the hardware must fetch massive, non-contiguous expert weight matrices from off-chip memory, saturating the available bandwidth and leaving compute units idle. This "memory wall" is the primary bottleneck of LLM inference performance on conventional compute-centric architectures, necessitating novel memory-centric solutions to mitigate the data movement overhead.

### *B. Memory-centric Accelerators*

The severe memory bottlenecks inherent to LLM inference have spurred research into specialized hardware accelerators. Initial efforts focused on software optimizations for existing GPU platforms. Systems including DeepSpeed-MoE and MegaBlocks employ techniques such as expert parallelism and optimized kernels to manage the massive memory footprint [17]. Such GPU/TPU serving stacks scale MoE by distributing experts across devices, which shifts the bottleneck to inter-device interconnect (NVLink) [17]. More fundamentally, they remain bound by the fixed bandwidth of off-chip DRAM and incur significant stalls during the memory-bound decode phase.

To move computation closer to the data, a wide variety of memory-centric accelerators have been proposed. Heterogeneous designs, including Atleus and H3D-Transformer, combine NVM PIM with high-endurance CMOS logic [9], [11]. However, these architectures are typically constrained by limited on-chip storage (MBs), forcing heavy reliance on off-chip DRAM for billion-parameter LLMs. Conversely, DRAM-PIM solutions offer capacity but incur high energy costs due to frequent refresh cycles [18]. Recent proposals stack 3D memories with PNM capability to resolve the storage requirement directly. For instance, Stratum proposes a system with tiered, monolithically-stacked 3D DRAM to provide massive, high-bandwidth capacity for expert weights [15]. Similarly, A3D-MoE and CHIME employ 3D heterogeneous integration, combining logic dies with DRAM to reduce data movement [8], [14]. These accelerators leverage the vertical data path between memory and computing cores to alleviate the memory bandwidth requirements of loading the expert parameters. However, these designs primarily utilize DRAM PNM and focus on the storage aspect, often adopting standard interconnects (e.g., mesh or ring). Here, the communication fabric is not explicitly tailored to handle non-deterministic traffic generated among computing cores as tokens are dispatched to spatially distributed experts.

In parallel, industry has begun exploring high-bandwidth NAND with near-memory compute for AI inference. SanDisk and SK Hynix have jointly defined the High-Bandwidth Flash (HBF) standard, which stacks NAND dies at HBM-comparable bandwidth with up to 512 GB per stack [19], [20]. SK Hynix's H³ architecture integrates HBM and HBF on a shared interposer alongside a GPU, reporting up to 2.69× higher throughput per watt over HBM-only configurations for LLM inference [21]. While these efforts confirm the practical feasibility of dense NAND for overcoming DRAM's capacity and refresh limitations, they rely on conventional Flash, which imposes high access latency and energy overheads. Moreover, similar to existing HBM-based PNM systems, they provide no intra-chiplet fabric to manage the irregular, scatter-gather data exchange among compute cores during MoE token routing.

Collectively, these efforts advance either storage density or raw bandwidth, yet none couples low-latency, high-density storage with a communication fabric designed for MoE traffic. This gap motivates a co-designed architecture that integrates such storage with a specialized NoC to break both the memory and communication bottlenecks of MoE inference.

## III. Problem Statement

Each layer within a transformer based LLM is composed of two computational modules: Multi-Head Attention (MHA) and Feed-Forward (FFN) [1]. While the MHA module remains consistent across LLMs and MoE models, the structure of the FFN is fundamentally different between these architectures. In MoE, the monolithic FFN of LLMs is replaced with a sparse layer of independent specialized experts (see Fig. 2) [2], [17]. Specifically, a MoE layer consists of a Router (or Gating Network) and a set of $N$ independent expert networks $E = \{E_1, \ldots, E_N\}$, each implemented as a standard FFN, where computation is inherently sparse, activating only a small subset of relevant experts per input [2], [17].

For a given input token representation $x$, the gating network $G(x)$ computes a sparse probability distribution, selecting only the top-$k$ experts (where $k \ll N$) to process the token, as shown in Fig. 2. Afterwards, the output $y$ is the weighted sum of the selected experts' outputs, as described in Eq. 1:

$$y = \sum_{i \in T} G(x)_i . E_i(x) \quad (1)$$

where $T$ represents the set of indices corresponding to the top-$k$ experts. While this sparsity keeps the number of active parameters per token comparable to a smaller model, the total number of model parameters to be stored for all experts can be orders of magnitude larger. This shift in neural architecture creates fundamental challenges: MoE models are computationally sparse but memory-dense, introducing significant irregularity in dataflow. The workload transitions from the uniform matrix multiplication characteristic of LLMs to scattered, input-dependent conditional computation in MoEs, creating both computation and communication bottlenecks.

First, the expert weights dominate the model size (often > 90%). Storing these massive parameters in off-chip Flash memory incurs prohibitive latency and energy penalties, while on-chip DRAM lacks the required density. Consequently, efficient expert computation demands a high-density storage medium capable of low-latency access.

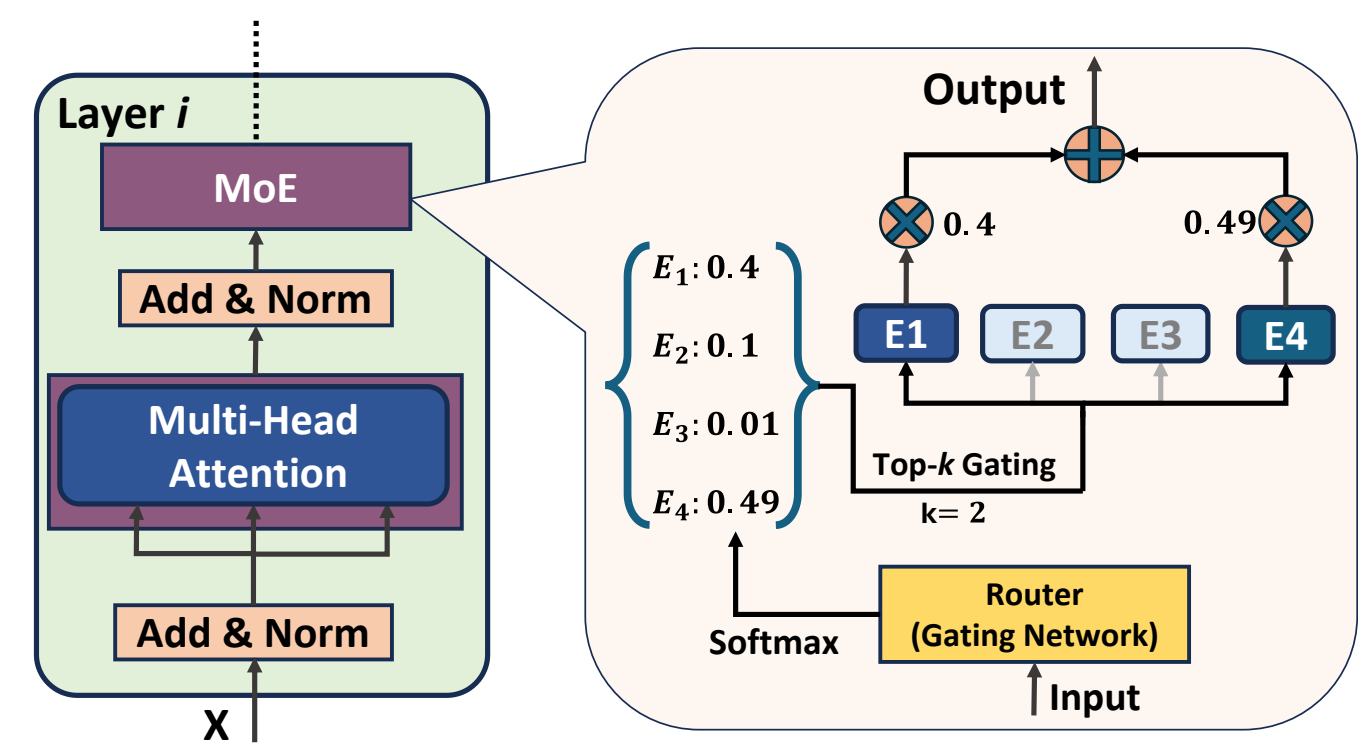

**Fig. 2:** High-level illustration of a MoE Layer with four experts per layer and top-$k$ ($k$=2) routing as an example. The model only selects the top $k$ experts with the highest gating probabilities per token within a layer.

Second, the active experts are dynamically allocated to the computing cores. Considering a manycore architecture, the gating mechanism dictates the workload distribution for a batch of input tokens. This results in highly variable, non-deterministic traffic patterns that change based on the semantic content of the batch [12]. Tokens must be dispatched to their assigned experts and output results aggregated, introducing significant network latency [13]. Furthermore, the aggregation of expert outputs imposes a synchronous aggregation operation, where the inference latency is dictated by the completion of the slowest expert.

From a combinatorial perspective, the allocation of $q$ cores to $e$ active experts can be modeled as partitioning the integer $q$ into $e$ parts. Consequently, the total number of unique NoC traffic scenarios ($S$), for a system with $q$ cores is given by the sum of partition functions, as expressed in Eq. 2:

$$S = \sum_{j=1}^{e} p_j(q) \quad (2)$$

Here, $p_j(q)$ represents the number of integer partitions of $q$ into exactly $j$ positive parts.

This combinatorial explosion of traffic patterns, combined with the memory-bound nature of experts, impose the following design challenges unique to MoE-based LLM inference:

1) **Memory Bottleneck of Experts:** The expert layers require massive storage density and high bandwidth to feed compute units. Existing memory hierarchies fail to balance the trade-off between density, leakage, and access latency for these sparse, static weights.
2) **Dynamic and Irregular Traffic:** Since expert selection is input-dependent, the specific source-destination pairs and traffic volumes cannot be determined prior to inference. Consequently, the routing of tokens from their source to the assigned expert computing cores for a particular batch creates a non-deterministic traffic pattern.
3) **Tail Latency:** The logical dependency on all selected experts (Eq. 1) imposes a synchronous gather operation. Thus, any network congestion or load imbalance affecting a single expert propagates to the entire batch.

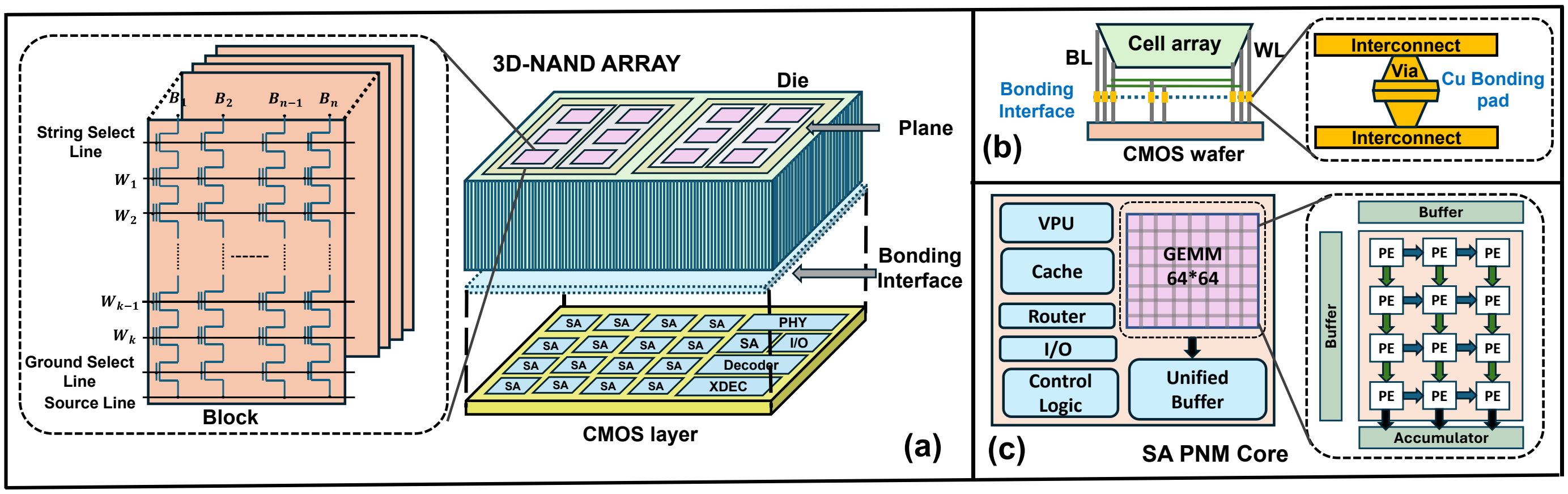

**Fig.3:** Illustration of the (a) NAND chiplet, (b) CMOS Directly Bonded to Array (CBA) technology used to integrate the 3D FeFET memory and the CMOS layer, and (c) systolic array chiplet. The dies are not drawn to scale. The figure is for illustrative purpose only.

Thus, the dominant unresolved bottlenecks in batched MoE inference arise from expert-side storage and the dynamic communication induced by sparse routing. Our goal is to precisely address this gap in the current state of knowledge.

## IV. ThAME: Heterogeneous Chiplet Architecture

In this section, we introduce ThAME, a three-dimensional heterogeneous architecture for MoE LLM acceleration. We first present the hardware substrate design and the co-design strategy that maps the MoE workload to address memory challenges. Subsequently, we elaborate on the specialized NoC backbone designed to mitigate the communication bottlenecks identified in Section III for the FeFET NAND base die.

### *A. ThAME Architecture*

The computational profile of MoE inference alternates between the compute-bound prefill and the memory-bound decode phase [4]. To address this phase-level heterogeneity, ThAME adopts a heterogeneous 2.5D system-in-package (SiP) design, integrating a compute-centric host chiplet with specialized memory-centric chiplets, as illustrated in Fig. 1.

**Memory-centric Chiplet:** ThAME utilizes 3D memory with PNM capabilities to address the memory-bandwidth bottleneck of the decode phase. However, a single memory technology cannot efficiently address the conflicting requirements of MoE workloads. The expert layers demand massive storage density for static weights, whereas the attention mechanism necessitates high endurance for frequent, write-intensive dynamic computation. Given the intrinsic physical limitations of memory devices, ThAME employs a heterogeneous memory substrate tailored to these distinct data access patterns.

*Expert Layers:* We propose *3D FeFET-NAND* memory to provide the requisite high-density storage for the expert computation. The NAND architecture offers superior storage density compared to DRAM by leveraging the mature, high-density, and cost-effective vertical stacking technology of Flash memory [22], [23]. The CMOS base die within the chiplet houses systolic array-based computing cores to support PNM and mitigate the latency of moving expert weights across the interposer. We detail the FeFET-NAND chiplet design in the next subsection.

*Attention Layers:* The attention mechanism generates dynamic intermediate states and requires frequent updates to the $KV$ cache. This workload involves write-intensive operations that would degrade the endurance of FeFET devices (typically limited to $10^8$-$10^{11}$ program/erase cycles) [24]. Consequently, we map these layers to *3D DRAM*-based PNM chiplets. Similar to the FeFET chiplets, the base die of the DRAM stack incorporates compute cores. We adopt a High-bandwidth Memory (HBM)-PNM architecture, following existing work to execute the dynamic read-modify-write operations required for attention computation [25]. Each attention head's KV matrices are assigned to a dedicated HBM die, and GEMV (General Matrix-Vector multiplication) units in the base die execute the attention score computation ($Score = Q \times K^T$) and the corresponding multiplication with the $V$ matrix ($Softmax(Score).V$) [25].

**Compute-centric Chiplet:** The prefill phase is compute-bound and exhibits high arithmetic intensity due to parallel token processing. Consequently, ThAME maps this phase to the host chiplet featuring a systolic array architecture (TPU V6e) optimized for dense matrix multiplications [26], [27]. The host also executes the gating network to determine the top-$k$ experts and dispatch tokens to the appropriate memory-centric chiplet.

Overall, the phase-aware mapping isolates the compute-intensive prefill kernel, while the memory-centric chiplets handle the decode phase. The heterogeneous chiplets are integrated on a passive silicon interposer, as illustrated in Fig. 1. Each of the memory chiplets is connected to the host chiplet via die-to-die links. The host chiplet is responsible for facilitating the data exchange among the 3D memory chiplets. Note that there is no direct communication path between the DRAM and FeFET chiplets. The interposer links are facilitated by the Universal Chiplet Interconnect Express (UCIe) standard, which is implemented via a physical layer (PHY) in each chiplet [28].

While UCIe manages the communication between chiplets, the non-deterministic traffic pattern generated by token-to-expert routing (as defined in Section III) requires a specialized intra-chiplet communication fabric. Therefore, the base dies of the FeFET chiplets are interconnected via a specialized NoC, optimized to handle the bursty and non-deterministic traffic inherent to batched MoE inference. In contrast, the deterministic dataflow of the attention mechanism allows the DRAM base dies to utilize a standard mesh interconnect.

### *B. FeFET NAND Chiplet*

The FeFET NAND architecture enables multi-GB on-chip storage capacity to store the expert parameters for each layer [29], [30], [31]. The 3D NAND array follows the Flash memory hierarchical organization consisting of channels, banks, planes, and blocks [29]. Fig. 3(a) presents the proposed FeFET NAND-based PNM chiplet. A single channel comprises multiple banks that support concurrent operations, with each bank further divided into planes to maximize internal parallelism. We map the expert weights such that each row of the weight matrix is stored within a single page (word line) of a memory block. This layout allows for the parallel retrieval of large weight vectors by activating a single word line (WL) [29], [31]. Importantly, FeFET technology operates at significantly lower voltages (~3V) compared to conventional Flash memory (~20V) [31]. This voltage reduction directly translates to lower power consumption and low-latency read throughput required to saturate the vertical connections to the base die, with FeFET achieving read speeds of around 5-10 ns (~ms for Flash) [31]. Further, recent 3D FeFET NAND architectures have shown excellent thermal stability and retention capabilities [24], [32].

**FeFET Compute Core:** The base logic die is the computational engine within the chiplet. The vertical integration of the base die is facilitated through CMOS Directly Bonded to Array (CBA) technology. CBA technology enables independent fabrication of both the memory cell and the CMOS layer in different technology nodes [33]. These wafers are bonded together using a Cu-to-Cu hybrid bonding process, as illustrated in Fig. 3(b) [33]. The vertical integration using hybrid bonding links enable massive vertical bandwidth, necessary to fetch the large expert parameters.

Each compute core features a systolic array consisting of processing elements (PEs) connected in a grid, a Vector Processing Unit (VPU) and local cache (see Fig. 3(c)). There are three primary strategies for mapping a computational kernel onto systolic arrays, namely, output stationary (OS), input stationary (IS), and weight stationary (WS) [34]. The "stationary" term determines which operand among the input, weight, or output remains fixed within the PE during the computation, directly governing the systolic array compute efficiency. Consequently, we evaluate these dataflows to identify the optimal mapping strategy for expert computations.

Furthermore, the number of cores is strictly constrained by the area footprint of the base die and the thermal budget of the chiplet. To prevent thermal throttling, the total power consumption ($P_{total}$) must satisfy:

$$P_{total} = P_{mem} + P_{peri} + P_{PHY} + q.P_{core} + P_{NoC} \leq P_{max} \quad (3)$$

where $P_{mem}$ is the power of the memory array, $q$ is the number of cores, $P_{core}$ is the power per core, $P_{NoC}$ is the NoC power. Additionally, $P_{peri}$ is the memory controller and peripheral circuitry power and $P_{PHY}$ is the D2D PHY power. Similarly, the logic area is constrained by the physical footprint of the memory base die ($A_{die}$). Along with compute cores, the CMOS base layer incorporates the peripheral circuitry and D2D PHY.

$$A_{TSV} + q.A_{core} + A_{PHY} + A_{peri} + A_{NoC} \leq A_{die} \quad (4)$$

Here, $A_{TSV}$ represents the area reserved for power delivery Through Silicon Vias (TSVs). Note that ThAME employs Cu-Cu hybrid bonding for high-speed data transfer between the memory cells and the logic die. Crucially, the hybrid bonding-based data I/O does not consume an active area in the logic die [33]. Based on these physical constraints, we determine the optimal core count $q$ and systolic array grid size, ensuring utilization of the vertical bandwidth provided by the hybrid bonding interface.

### *C. FeFET NAND NoC Design*

The FeFET NAND chiplet hosts a set of homogeneous compute cores interconnected via the NoC in the base die. As detailed in Section III, the NoC design must be robust against dynamic non-deterministic MoE traffic. Each candidate topology must be evaluated against the complete theoretical possible space of Tensor Parallelism ($TP$) and Expert Parallelism ($EP$) configurations. The core design question is therefore: given a fixed system size (number of cores) and *a priori* unknown traffic pattern, how do we design a NoC that simultaneously minimizes congestion, load imbalance, and hardware cost (area).

For each layer, the input activations are received at the PHY of the FeFET NAND chiplet from the host chiplet. The per-token routing gives rise to an irregular, scatter-gather traffic pattern. The activations must be routed from the host to the selected experts (a one-to-many pattern), and the computed outputs from selected experts must be aggregated back to the host (a many-to-one pattern). During batched inference, the gating network's token allocation can be highly skewed. For instance, experts receiving more tokens require their weight matrices to be partitioned across multiple cores to accelerate computation ($TP > 1$), while an expert receiving a small number of tokens can be mapped to a single core ($TP = 1$). As established in Section III (Eq. 1), this logical dependency on all selected experts imposes a synchronous gather operation. Thus, any network congestion affecting a single expert propagates to the entire batch, increasing tail latency and degrading throughput.

**Multi-Objective Optimization (MOO) Formulation:** The MoE token routing, characterized by a one-to-many routing from the host to the compute cores, followed by a many-to-one aggregation to the host, makes a tree-based NoC a natural architectural fit [35]. However, connecting individual cores directly to a monolithic tree NoC introduces scalability issues and congestion. As an example, when an expert is mapped across multiple cores using $TP$, these cores need to exchange activations and partial sums. Routing this intra-expert traffic through a tree topology would necessitate multi-hop paths for the latency-sensitive $TP$ computation.

To resolve this communication challenge, we propose a hierarchical NoC design, as illustrated in Fig. 4. At the lowest level, cores are grouped together and interconnected via local crossbars. These crossbars provide non-blocking and fast communication, confining the activations and partial-sum traffic mostly within the local crossbars [36]. A tree topology is employed as the upper-level communication backbone to handle inter-crossbar traffic and activations to and from the host. This upper-level network starts below the PHY. Moreover, the NoC needs to account for transferring expert parameters from the FeFET memory tiers to the compute cores. Thus, the level of the tree where the memory interface injects traffic becomes crucial. For example, if the memory-interface points are at the root of the tree, the volume of the weight traffic can completely saturate the tree links. Conversely, attaching the memory interface directly at the individual cores eliminates the bottleneck but incurs a massive area overhead.

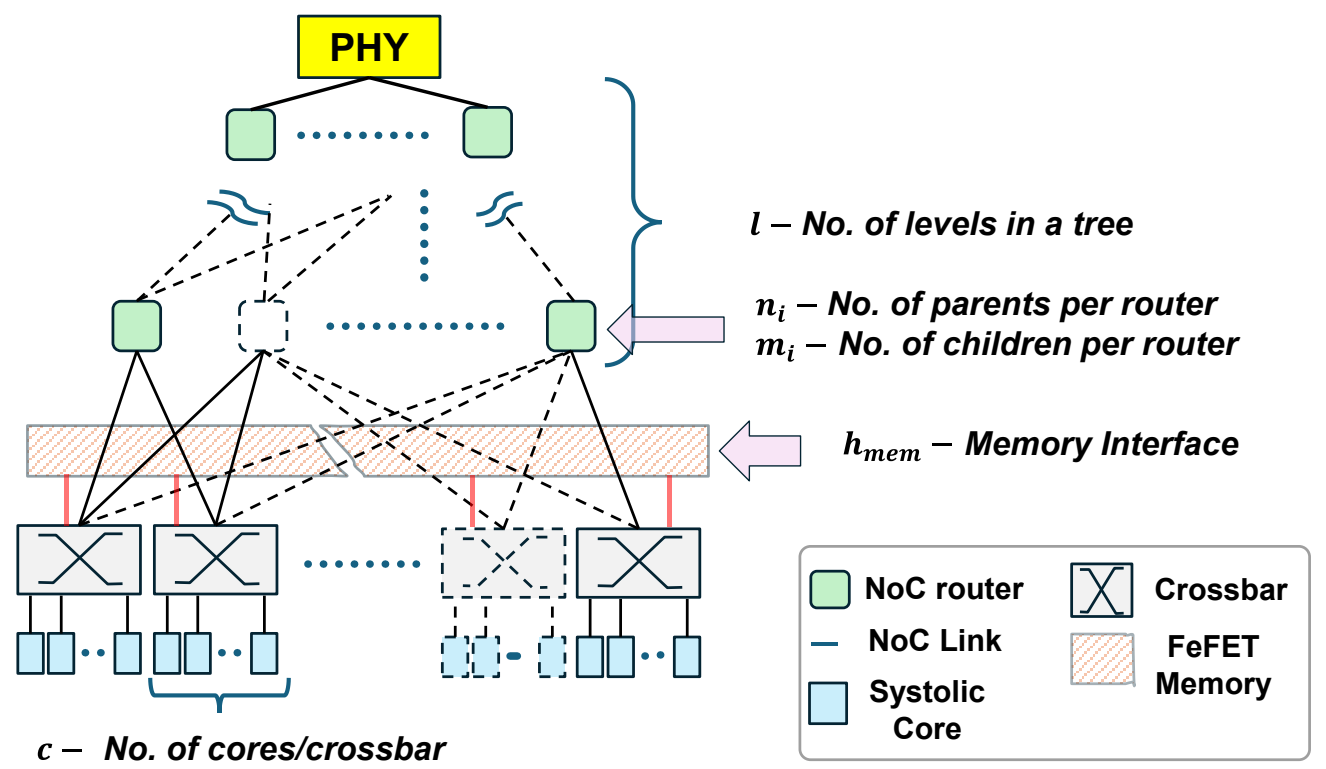

**Fig.4:** NoC design variables for MOO problem: $\mathbf{d} = (c, l, \boldsymbol{m}, \boldsymbol{n}, h_{\text{mem}})$, where the design variables includes the crossbar size $(c)$, tree depth $(l)$, vectors $(\boldsymbol{m})$ and $(\boldsymbol{n})$ represent the number of children and the number of parents per router assigned at each level of the tree, and memory injection level $(h_{\text{mem}})$.

We formulate this NoC design as a MOO problem to determine the optimal number of cores per crossbar, the number of levels of the tree, the number of children and parents per router at each level of the tree, and the memory interface point. Each candidate NoC design $d$ corresponds to a specific configuration of this hierarchical tree in the explored design space $D$ during optimization. We describe the key elements of our MOO framework as follows:

1) **Design Variables:** $d = (c, l, \boldsymbol{m}, \boldsymbol{n}, h_{mem})$, where $(c)$ is the size of the local crossbars (the number of cores per crossbar), $(l)$ is the number of tree levels (see Fig. 4). The vectors $(\boldsymbol{m})$ and $(\boldsymbol{n})$ represent the number of children and the number of parents per router assigned at each level of the tree, respectively. The parameter $(h_{mem})$ is the level of the NoC at which the memory injects traffic.
2) **Constraints:** First, the NoC must guarantee full host-to-core connectivity: there must be at least one communication path between the host and all cores. Second, the number of links must not exceed that of a standard 2D mesh NoC.
3) **Objectives:** Our goal is to "robustly" maximize NoC performance while minimizing area across all combinatorial space of traffic patterns.

Conventional NoC frameworks optimize architectures for a deterministic compute mapping [36]. Since the MoE traffic pattern is entirely input-dependent, the traffic pattern is *a priori* unknown. Our key novelty in the NoC design lies in optimization for the predicted behavior across the combinatorial space of traffic scenarios $(S)$. Note that predicting these exact runtime traffic patterns is not practical as it would stall inference. Consequently, the optimization assumes no prior knowledge of the workload, or the token distribution among experts, treating each traffic pattern as equally probable. This uniform prior is a design decision in our optimization. The NoC topology is fixed at fabrication and must serve arbitrary models, datasets, and routing behaviors over the lifetime of the chip, none of which are known a priori. Therefore, optimizing under a uniform prior yields a worst-case robustness guarantee against any future per-expert core allocation within the base die due to the gating network's token-to-expert distribution.

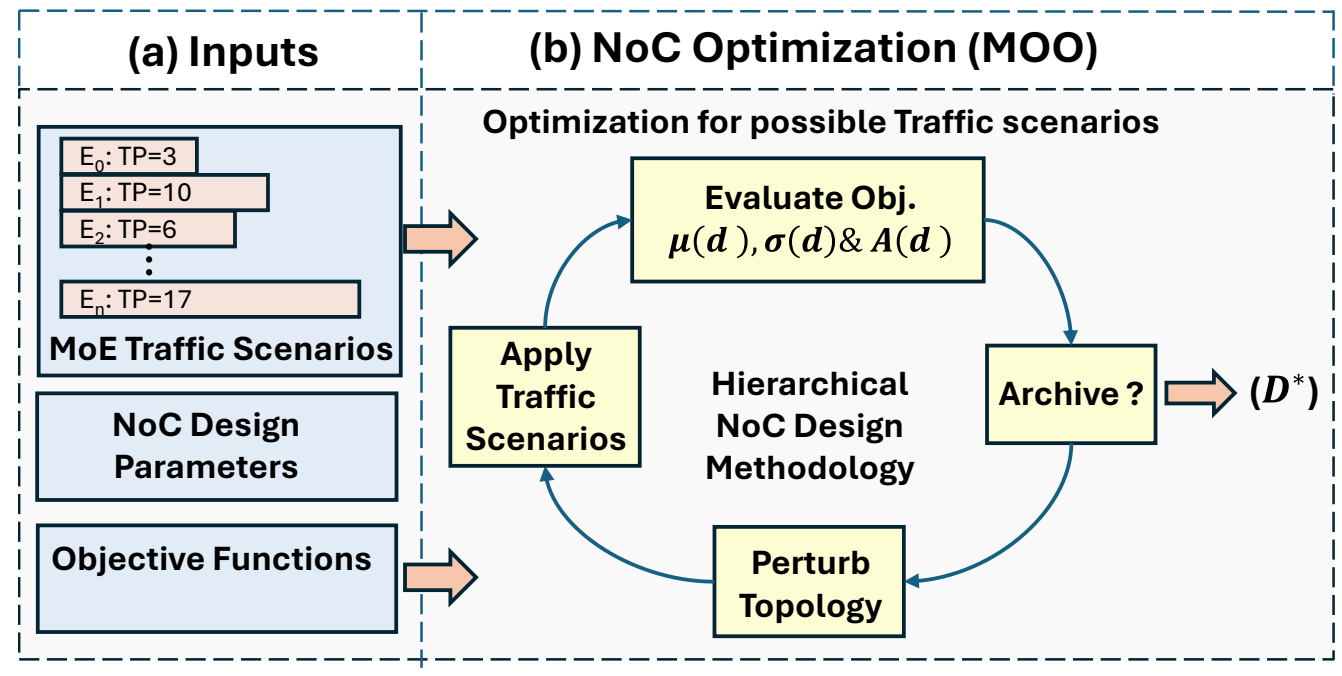

**Fig.5:** Workflow of the ThAME's NoC optimization methodology, showing (a) input for MOO, and (b) High-level MOO solver.

We therefore optimize three objectives across this space: mean link utilization $\mu(d)$, standard deviation of the traffic utilization $\sigma(d)$, and the total NoC area $A(d)$ of each design $(d)$. Minimizing $\mu(d)$ ensures that the NoC maintains sufficient global bandwidth to prevent congestion when an input triggers a large number of tokens to many experts. Furthermore, the MoE layer requires a synchronous gather operation once all the experts have computed their outputs, a high variance in link utilization would cause localized congestion. This imbalance delays individual experts, creating a tail-latency stall for the entire batch. Thus, minimizing $\sigma(d)$ guarantees that the tail-latency stall is reduced.

For a given design $(d)$ and a traffic scenario, $(s \in S)$, let $F_{ij}(s)$ denote the volume of traffic between routers $(i)$ and $(j)$. The link utilization on a specific link $(K)$ is defined as:

$$u_k(s) = \sum_{i=1}^{L} \sum_{j=1}^{L} F_{ij}(s) q_{ijK} \tag{5}$$

where $L$ is the total number of routers, and $q_{ijk}$ is a Boolean indicator if link $(K)$ is being utilized between routers $(i)$ and $(j)$. For a single traffic scenario $(s)$, the mean $\mu(d,s)$ and standard deviation $\sigma(d,s)$ of link utilization can be represented as:

$$\mu(d,s) = \frac{1}{P} \sum_{K=1}^{P} u_K(s) \tag{6}$$

$$\sigma(d,s) = \sqrt{\frac{1}{P} \sum_{K=1}^{P} \left(u_K(s) - \mu(d,s)\right)^2} \tag{7}$$

where $(P)$ is the total number of links in the design $(d)$ [37]. To guarantee robustness against any dynamic MoE routing, we compute the average mean link utilization and standard deviation across all traffic scenarios $(S)$:

$$\mu(d) = \frac{1}{|S|} \sum_{s \in S} \mu(d,s), \sigma(d) = \frac{1}{|S|} \sum_{s \in S} \sigma(d,s) \tag{8}$$

Finally, to quantify the hardware cost $A(d)$, we model the total NoC area as follows:

$$A(d) = \sum_{i \in R_{tree}} A_{router}(i) + \sum_{j \in R_{cb}} A_{cb}(j) + A_{mem} \tag{9}$$

where $R_{tree}$ is the set of routers in the tree, $R_{cb}$ is the set of local core crossbars, $A_{router}(i)$ and $A_{cb}(j)$ denote the area of router $i$ and crossbar $j$ respectively, and $A_{mem}$ represents the area of the memory interface.

By jointly optimizing these objectives, we obtain NoC designs that are inherently robust to the variability and skew present across all traffic patterns. We represent the MOO formulation as follows:

$$\mathcal{D}^* = MOO\big(\mu(d), \sigma(d), A(d)\big) \tag{10}$$

where ($\mathcal{D}^* \subseteq D$) is the set of pareto-optimal designs, as illustrated in Fig. 5. A design is considered Pareto-optimal if no single objective can be further improved without degrading others. We can employ MOO solvers such as AMOSA to solve this MOO problem [38]. Subsequently, we perform cycle-accurate simulations for each design ($d$) in ($\mathcal{D}^*$) to identify a NoC design that provides a suitable trade-off (shown in Fig. 7 later) between performance and area overhead.

*D. End-to-End Pipeline for ThAME*

ThAME implements a pipelined dataflow that aligns the heterogeneous computational kernels with their optimal hardware substrates. Fig. 6 illustrates the overall execution flow of the ThAME framework. For each batch, execution begins at the host chiplet with the prefill phase computation to populate the $KV$ cache ❶. Subsequently, the system transitions into the autoregressive decode phase where the host dispatches the intermediate token representations (queries) to the 3D DRAM-PNM chiplets via the interposer. The DRAM base-die compute cores execute the attention mechanism locally, utilizing the intra-chiplet bandwidth to perform the requisite read-modify-write operations on the $KV$ cache ❷. This execution confines the $KV$ cache within each DRAM chiplet, eliminating inter-chiplet data movement. The resulting attention activations are returned to the host chiplet, which evaluates the gating network to determine the top-$k$ expert assignments ❸.

The gating network is a single projection mapping each $d_{model}$ dimension activation to $N_{experts}$ routing logits, where $d_{model}$ and $N_{experts}$ denote the model embedding dimension and the total number of experts within a layer, respectively. Each logit is an unnormalized score measuring the affinity of the token for a given expert. A softmax over these logits yields the routing weights $G(x)$, from which the top-$k$ highest-scoring experts are selected (Eq. 1). This computation involves a negligible amount of compute operations compared to the expert layer. Specifically, the gating network maps the token representation to the expert domain, requiring $O\big(d_{model}.N_{experts}\big)$ operations per token. In contrast, the top-$k$ expert computation requires $O\big(d_{model}.d_{ff}.k\big)$ operations, where $k$ and $d_{ff}$ denote the active experts and the feed-forward hidden dimensions, respectively. Since $N_{experts}$ is substantially smaller than the activated expert's hidden dimension $d_{ff}$, the gating overhead is negligible relative to the overall MoE execution time.

The gating network fixes the compute workload of each active expert through its token assignment. Following this stage, the host chiplet utilizes a scheduler to assign computational cores to the active experts such that all experts complete at approximately the same time ❹. Let $e$ active experts be selected by the gating network for a given batch and let $L_i(t_i)$ denote the compute latency of expert $i$ when allocated $t_i$ cores. Due to the tiling behavior of the systolic array, $L_i(t_i)$ is a non-linear, staircase function of the core count. For our scheduling strategy, we formulate core assignment as an integer min-max optimization:

$$\min_{\{t_i\}} \max_{i \in \{1,\ldots,e\}} L_i(t_i) \text{ s.t. } \sum_{i=1}^{e} t_i \leq q \tag{11}$$

where $q$ is the total number of available cores. To solve Eq. 11, we implement a hardware unit within the host chiplet. It performs a binary search over the expert compute latency domain $[C_{min}, C_{max}]$, where ($C_{min}$) denotes the fastest compute latency where all the tokens are equally distributed on the computing cores under uniform top-k distribution and ($C_{max}$) denotes the worst-case single-core expert latency. Since $L_i(t_i)$ is a monotonic function, the search converges in $\lceil log_2(C_{max} - C_{min})\rceil$ iterations. We later demonstrate that the scheduler takes a small number of cycles and is negligible relative to the expert compute and communication phases. The resulting core allocations per expert vary combinatorially across batches as the gating network's token distribution changes with each input. Each distinct allocation produces a unique spatial traffic pattern within the base die (see Section III, Eq. 2). Consequently, our hierarchical NoC design (Section IV-C) is critical to dynamically route these tokens, ensuring that the scheduler's compute load-balancing is not undermined by network contention ❺. Finally, the computed expert outputs are aggregated back to the host via the synchronous gather operation ❻ (see Fig. 6).

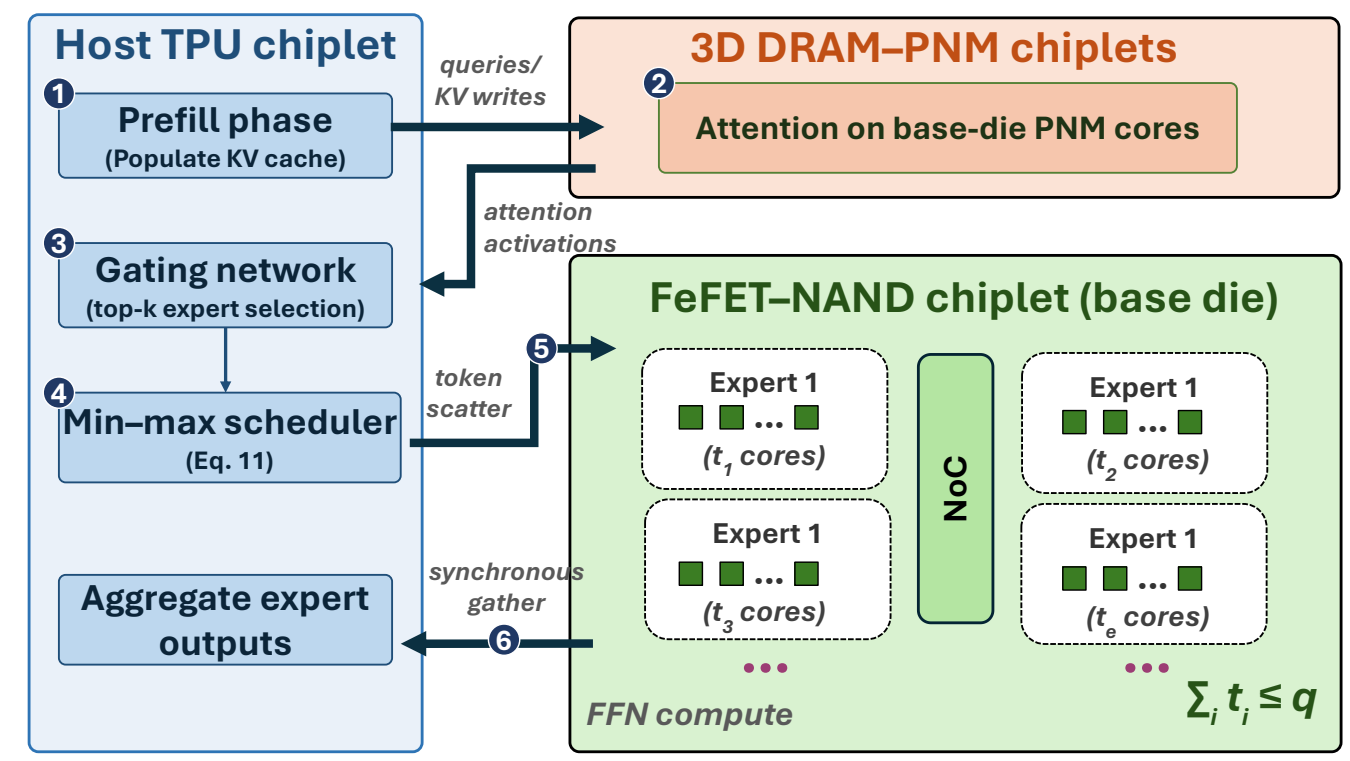

**Fig. 6:** End-to-end ThAME execution pipeline highlighting the prefill and decode phase mapping for MoE serving. The chiplets are not drawn to scale.

Overall, this execution pipeline mitigates the inter-chiplet communication bottleneck. In conventional architectures, processing a single token during decode necessitates fetching massive, expert weight matrices $O(d_{model}.d_{ff})$ across the memory bus. ThAME, conversely, leverages memory-centric compute to ensure that these expert weights remain localized within the high-density FeFET memory. As shown in Fig. 6, the only data traversing the 2.5D interposer links are the intermediate token activations. For a batch size of $B$, this inter-chiplet traffic is bounded to $O(B.d_{model})$, representing an activation vector footprint that is orders of magnitude smaller than the expert parameters.

## V. Experimental Results

In this section, we first describe the experimental setup used for the performance evaluation of ThAME. Next, we analyze the ThAME compute core design followed by the NoC design within the FeFET-NAND die. Afterwards, we analyze the effectiveness of both the computation and communication optimization during expert processing. We then present a sensitivity analysis under degraded FeFET read latency. Finally, we present a comparative and end-to-end analysis of ThAME against the baselines in terms of performance and energy during MoE LLM inference.

### *A. Experimental Setup*

We consider a diverse range of LLM models in our evaluation, as outlined in Table I. These models represent the typical size and diversity of real-world use cases for deployment. A 16-bit fixed-point data representation was used as the precision of all model parameters and activations.

**Compute Modeling:** We employed a cycle-accurate simulator to determine the latency and energy for each compute unit. *SCALE-Sim* was used to model the latency for each systolic array within the base die of memory chiplets and the host [34]. For the FeFET-NAND PNM chiplet, a modified version of *NeuroSim* was employed to capture the latency and energy of the 3D memory array [39]. NeuroSim has been validated against post-layout SPICE simulations of a taped-out PIM macro, showing chip-level prediction error below 1% after calibration for area, latency, and energy metrics. Similarly, SCALE-Sim has been verified against production TPU hardware with median relative error below 3% for representative operators [34]. Table II provides the detailed specifications of the FeFET-NAND, DRAM, and host chiplets used in our evaluation.

**Communication Modeling:** BookSim2, a cycle-accurate simulator, was used to model the intra-chiplet NoC [40]. The inputs to the *BookSim2* are the connectivity between routers, the intra-chiplet traffic, the link bandwidth, and the routing algorithm. Table III shows the NoC parameters for modeling the inter-core links within the FeFET-NAND chiplet.

In this work, we have implemented ThAME using a configuration with two FeFET and two DRAM memory chiplets along with a systolic array (TPU v6e) based host chiplet on a silicon interposer. Each of the memory chiplets is connected to the host chiplet in a point-to-point manner through the interposer. Note that this specific configuration does not affect the generalizability of our findings. We compare ThAME against two heterogeneous accelerators, Stratum (a DRAM-PNM and compute-centric host) and H3D-T (a 3D-stacked FeFET, SRAM, and TPU architecture) [9], [15]. The base die area for both the DRAM and the FeFET-NAND chiplet is kept consistent at 121 mm$^2$, similar to the baseline Stratum for comparison. For H3D-T, we instantiate multiple instances to establish an iso-area configuration for a fair evaluation.

The base die within the FeFET-NAND chiplet consists of systolic array-based computing cores. We follow existing work to determine the area and power for the memory peripherals, PHY and the power-delivery network for the base die [15]. To perform the thermal evaluation and to ensure the feasibility of the 3D-stacked base die without inducing thermal throttling, we employed HotSpot [41]. We performed a steady-state simulation at a 45 C ambient to establish an allowable budget of 50W for the base die, using the parameters detailed in Table III. Given the area budget and FeFET memory bandwidth in our experimental setup, we provisioned a total of 4,096 PEs per core in a 32-core system under maximum utilization where all cores are simultaneously active. Table IV details the specifications of the systolic array cores. We employ the Archived Multi-Objective Simulated Annealing (AMOSA) algorithm to identify the Pareto-Optimal set of designs ($\mathcal{D}^*$) for our hierarchical NoC design. AMOSA is a simulated annealing-based MOO solver in which the optimization process is guided using an archive of previously explored solutions [38]. The design space exploration and evaluation were performed on an Intel Core i5 CPU @2.4GHz machine with 16 GB RAM in which convergence was achieved in approximately 30 minutes.

TABLE I
GENERATIVE/LLM MODELS AND SPECIFICATIONS

| Generative Models | Params | Type | Specification [layers, d_model, d_ff] | Expers: Active/ token (k) | Experts: Total |
|---|---|---|---|---|---|
| Deepseek [47] | 16B | MoE | [28, 2048,1408] | 6 | 64 |
| Olmoe1b [49] | 7B | MoE | [16, 2048,2048] | 8 | 64 |
| Qwen1.5 [50] | 2.7B | MoE | [24, 2048,1408] | 4 | 60 |
| Qwen3 [48] | 30B | MoE | [48, 2048,768] | 8 | 128 |
| Llama3 [46] | 8B | Llm | [32, 4096, 14336] | NA | NA |

TABLE II
ARCHITECTURE SPECIFICATIONS

| **FeFET Chiplet:** | |
|---|---|
| 3D NAND Array | 8 Channels, 4 Banks, 4 Planes, 64 Blocks<br>WL – 128, BL – 9216 |
| NAND Cell [31] | WL Voltage (Program/Inhibit/Read) – 3V/1.7V/0.5V, BL Voltage (Program/Inhibit/Read) – 3V/1.7V/0.5V, Single-level cell (SLC) device, read latency – 10 ns, Tech node – 32 nm, 121 mm$^2$ |
| **Host Chiplet** | |
| TPU v6e [26] | Systolic Array (256×256 PEs) - 8 , VPU – 8, Total SRAM – 256 Mb, 2 GHz clock, TSMC's 5nm process node |
| **DRAM Chiplet** | |
| HBM-PNM [25] | 8-HI, 2 ranks per stack,8 pseudo-channels per die, 1z-nm (10 nm class) DRAM process node, 7 nm CMOS base die process node, 121 mm$^2$ , Interface - 1024-bit bus per stack at 5.2 Gbps per pin |

TABLE III
DESIGN PARAMETERS

| | |
|---|---|
| UCIe [28] | NoP Frequency – 2 GHz, Energy per bit – 0.5 pJ |
| NoC [40] | Bandwidth – 32 GB/s, Link width – 64, Mesh- XY routing, Tree- LCA, Ring/Torus - Dimension order |
| Thermal Modeling | Vapor chamber heat sink, Thermal conductivity - 5000 W/(m·K), Heat Capacity of $10^6$ J/(m³·K), $T_{peak} - 360K$, Convection resistance - 0.05 K/W [15] |

TABLE IV
FEFET BASE DIE PARAMETERS

| 32 compute cores per chiplet | |
|---|---|
| Core | 64×64 array of PEs, 1 VPU , Total SRAM – 1 Mb, 800 MHz clock, 1.13 W, 2.05 mm$^2$, Tech node – 7 nm |
| Hybrid Bonding | Pitch – 5 $\mu$m [43], Bandwidth – 5 TB/s |

TABLE V
DATAFLOW COMPARISON WITH LLAMA3

| Dataflow | Inner-Loop Dimension | Per-Tile Overhead | Total Cycles (Norm) | Bandwidth (TB/s) |
|---|---|---|---|---|
| OS | K (4096/14336) | 3.0% | 1× | 3.28 |
| WS | M (32) | 85.6% | 3.39× | 0.945 |

### B. ThAME Compute Core

Each expert comprises three compute operations: gate, up, and down projections [2]. In standard GEMM notation, each expert projection computes $\text{Out}[M,N] = \text{Act}[M,K] \times \text{W}[K,N]$, where $M$ corresponds to the batch size $B$, $K$ denotes the reduction dimension, and $N$ is the output dimension. During the decode phase, these GEMM operations exhibit a pronounced asymmetry: $M$ is small ($B \in \{8, \dots, 64\}$) while $K$ and $N$ are large (e.g., $d_{model} = 4096$, $d_{ff} = 14336$). The dataflow within the systolic array determines the inner-loop (stationary dimension) during computation, which in turn affects the data reuse [34]. During tiling, the activations are accumulated over the inner-loop with fill/drain necessary after each iteration. Note that fill/drain cost is dataflow-dependent: OS incurs $R + C - 2$ cycles per tile, while WS incurs $R + 2C - 2$ cycles per tile, where $R$ and $C$ denote the row and column dimensions of the systolic array, respectively [34].

The $OS$ (output stationary) dataflow places the large $K$ dimension (4096 for gate/up, 14336 for down) in the inner loop, whereas $WS$ (weight stationary) is constrained to stream the $M$ dimension ($B = 32$), which is significantly smaller. For a 64×64 array, an inner loop of $K = 4096$ yields a per-tile latency overhead of $127/4223 = 3.0\%$. In contrast, $WS$ has an inner loop of $M = 32$ that leads to a per-tile overhead of $190/222 = 85.6\%$ in idle cycles. Given $K \gg M$ for all three GEMM operations during the decode phase, the $OS$ dataflow exhibits significantly superior compute efficiency compared to $WS$. Table V presents a comparative analysis between the two dataflows for the Llama model as an example. We do not consider $IS$ (input stationary) since it requires an additional adder tree per row to sum partial products each cycle [42]. As seen in Table V, $OS$ leads to a higher throughput, although with a higher bandwidth requirement. Crucially, we harness the 3D FeFET-NAND memory, which provides $\sim$5TB/s of vertical bandwidth (see Table IV), to deliver the requisite sustained throughput to support the $OS$ dataflow [43], [44].

### C. ThAME NoC Optimization

This section evaluates the performance benefits of the proposed hierarchical NoC. As detailed in Section IV.C, we must evaluate the Pareto-optimal NoC designs ($\mathcal{D}^*$) across all traffic scenarios ($S$). Following Eq. 2, we determine the unique traffic possibilities for our system within the base die comprising 32 cores. While this formulation defines the complete theoretical traffic space, a vast majority of these possibilities represent inefficient hardware mappings. Allowing an arbitrary odd number of systolic array cores to an expert leads to inefficient hardware utilization since matrix tiling necessitates an even division in the number of tiles. Thus, we prune the possibility space to a hardware-valid subset denoted by $S_{valid} \subset \{1\} \cup \{2,4,6, \dots, U\}$, comprised of even integers along with the unpartitioned, single-core expert mapping. This pruned set captures the realistic, spatial traffic distributions that the network may experience during input-dependent batched inference. Fig. 7 shows the resulting Pareto front, plotting average latency against NoC area for $S_{valid}$ considering DeepSeek MoE as an example. Note that minimizing the mean $\mu(d)$ and standard deviation $\sigma(d)$ of link utilization reduces network congestion and prevents tail-latency stalls, which directly translates to minimizing NoC latency.

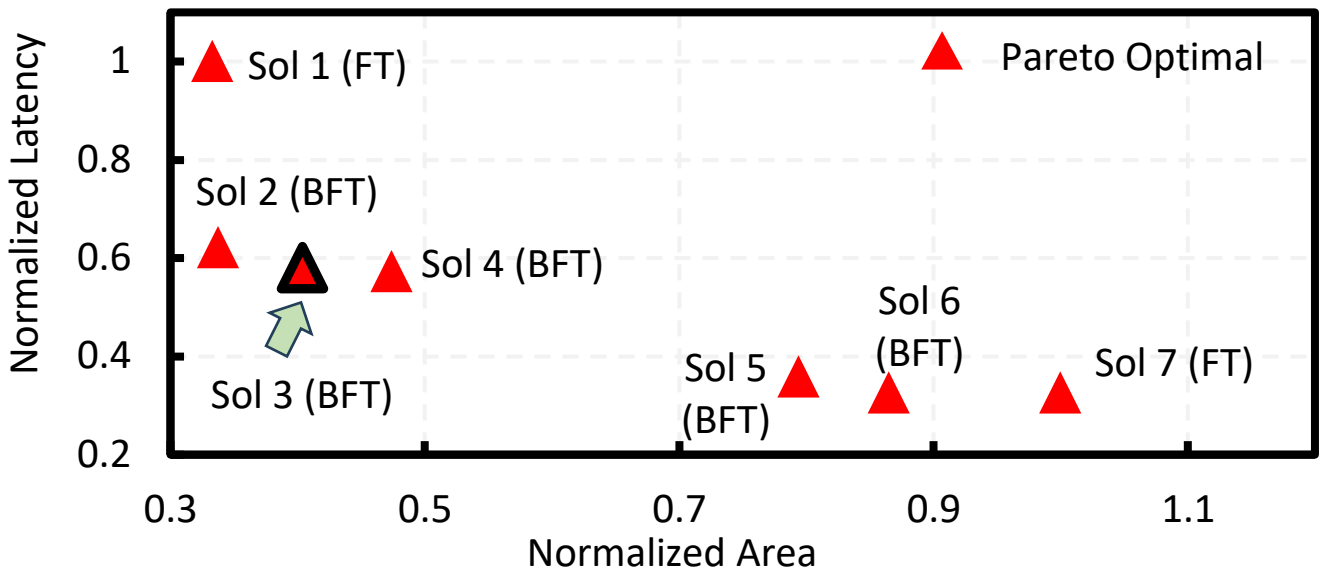

**Fig. 7:** Pareto-front of candidate NoCs via cycle accurate simulations showing normalized average communication latency versus normalized NoC area.

Given the strict bandwidth requirements from the PHY to the cores, we observe that specific configurations of Fat Tree (FT) and Butterfly Fat Tree (BFT) are viable candidates as the upper-level (below PHY) communication backbone (see Fig. 4). In Fig. 7, each design in ($\mathcal{D}^*$) is categorized by its topology: FT and BFT. We observe a clear trade-off between area and latency. Solutions #6 and #7 achieve the lowest latency at the cost of area, while solutions #1 and #2 minimize area but suffer from high mean latency. Solution #3 offers the best compromise, achieving 43% reduction in latency compared to solution #1 at the cost of 18% more area. Consequently, we choose Solution #3 as the NoC backbone, as it provides the most effective balance between latency and silicon footprint.

Solution #3 corresponds to a BFT(2,4,2) with design parameters: $d = (c{:}\,4, l{:}\,2, \boldsymbol{m}{:}\,[\mathbf{4},\mathbf{4}], \boldsymbol{n}{:}\,[\mathbf{2},\mathbf{2}], h_{mem}{:}\,0)$ for our hierarchical NoC. Specifically, this configuration groups $c = 4$ cores per local crossbar and employs an $l = 2$ level upper tree (see Fig. 4). At each level, the routers are connected to four children and two parents. Furthermore, the memory interface is injected at the lowest level of the tree ($h_{mem} = 0$) to handle the massive expert weight traffic without saturating the upper-level communication backbone.

We compare this MOO-optimized hierarchical NoC topology, using Least-Common-Ancestor (LCA) routing, against three NoC baselines within the base die of the FeFET NAND chiplet: a 2D Mesh using XY routing, and Torus and Ring topologies using Dimension-Order routing. Fig. 8 presents the normalized latency and area for each topology, evaluated across traffic scenarios ($S_{valid}$). The BFT(2,4,2) provides a $4.1\times$ reduction in average latency over the Ring topology. The hierarchical tree structure confines intra-expert partial-sum and activation traffic mostly within local crossbars, leaving the tree dedicated to the scatter-gather traffic patterns intrinsic to MoE token routing. In contrast, the Mesh and Torus NoC designs distribute links uniformly across all routers. The Torus gives marginal latency benefit under non-uniform, input-dependent MoE traffic over Mesh, but at the cost of 48% more area. Conversely, the Ring topology, despite having the smallest area, suffers from a high average latency due to multi-hop data exchange.

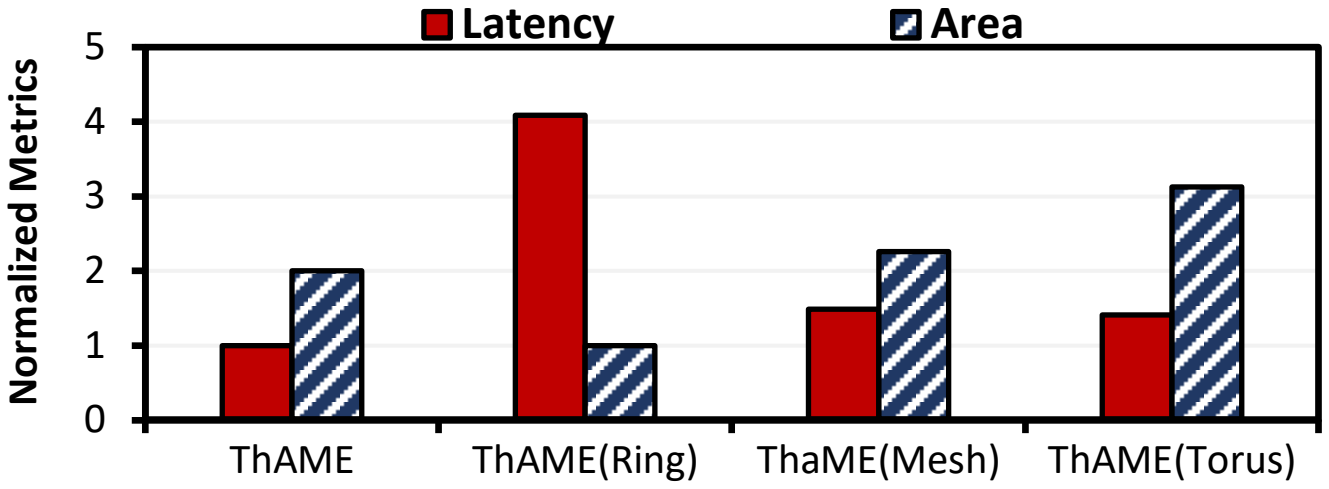

**Fig. 8:** Comparative analysis of normalized communication latency and NoC area for ThAME across baseline topologies (Ring, Mesh and Torus).

### *D. Expert Serving and NoC Evaluations*

The expert computation within the FeFET NAND base die necessitates a strict balance between computation and communication to achieve high throughput. As established in Section III, the inference latency of the entire batch is dictated by the slowest core. We evaluate the effectiveness of our hierarchical NoC design (Solution #3 in Fig. 7) by analyzing the compute-communication mismatch, defined as, $|Comm - Comp|/Comp$ across the entire theoretical combinatorial space of traffic patterns ($S_{valid}$). Fig. 9 presents the Cumulative Distribution Function (CDF) of this mismatch metric for the Qwen1.5 MoE model. As illustrated in Fig. 9, ThAME successfully mitigates the network-side tail-latency bottleneck, maintaining less than a 2% mismatch for over 60% of the valid spatial traffic configurations while enforcing a worst-case maximum deviation of 27%. ThAME's NoC systematically bounds the link utilization variance $\sigma(d)$, effectively minimizing performance deviations across the disparate, input-dependent batch routing scenarios. Conversely, conventional Mesh and Torus topologies impose significant network contention, exhibiting a 13% mismatch at the 60$^{th}$ percentile and degrading to a deviation of 35% for the 90$^{th}$ percentile. We exclude the Ring topology from this plot as its bandwidth limitations induce a communication bottleneck, rendering at least a 300% mismatch.

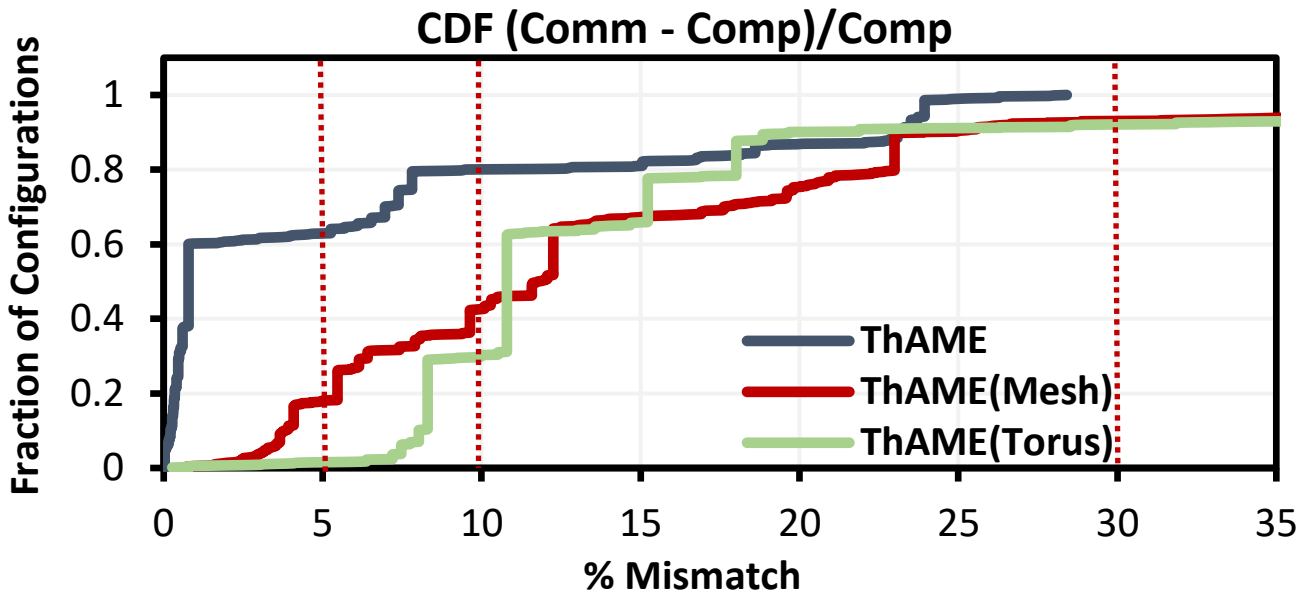

**Fig. 9:** Cumulative Distribution Function (CDF) of the compute-communication latency $|Comm - Comp|/Comp$ across $S_{valid}$ for the Qwen1.5 model.

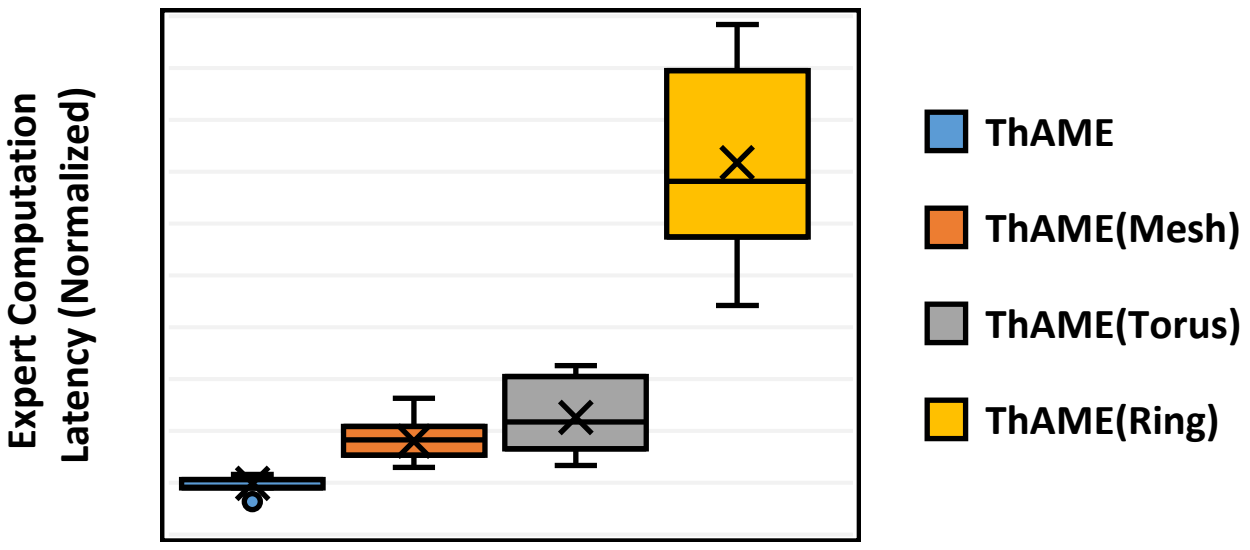

**Fig. 10:** Box-and-whisker distribution of normalized end-to-end expert latency for the Qwen1.5 MoE model evaluated on the MMLU dataset ($B = 32$).

The preceding analysis corroborates the theoretical robustness of the NoC against all valid spatial partitions. We now validate our optimization utilizing a real-world deployment scenario to determine the exact token-to-expert distributions. Specifically, we evaluate the end-to-end expert processing latency using the Massive Multitask Language Understanding (MMLU) dataset with a batch size of 32 ($B = 32$) as an example. The MMLU dataset provides a highly diverse set of multiple-choice questions spanning 57 distinct subjects (ranging from STEM to humanities). The workload traces for the simulation were generated using PyTorch to capture the non-deterministic MoE traffic.

Fig. 10 presents a box-and-whisker distribution of the normalized latency across the complete dataset. Here, the compute delay cycles are kept consistent across all NoC configurations to isolate the communication efficiency. As shown in Fig. 10, ThAME outperforms the baseline interconnects, providing superior median performance and significantly tighter latency bounds. Interconnects such as Mesh and Torus fail to accommodate the non-deterministic scatter-gather traffic, causing significant network communication overhead. Notably, the worst-case traffic tail latency in our hierarchical NoC is better than the best-case latency observed in both Mesh and Torus configurations. This corroborates the hypothesis that tail-latency advantage established over the full combinatorial space in Fig. 9 holds under the skewed traffic that occurs for a real workload.

**Ablation Analysis:** To isolate the contribution of each architectural component in ThAME, we evaluate four distinct configurations with a batch size of 32 for the Qwen1.5 and DeepSeek models: (1) ThAME, (2) ThAME-Mesh (FeFET-NAND + Mesh) to isolate the NoC contribution, (3) Stratum-Mesh (DRAM-PNM + Mesh) to isolate the memory substrate contribution, and (4) the original Stratum baseline. Fig. 11 shows the normalized decode latency and energy for each configuration.

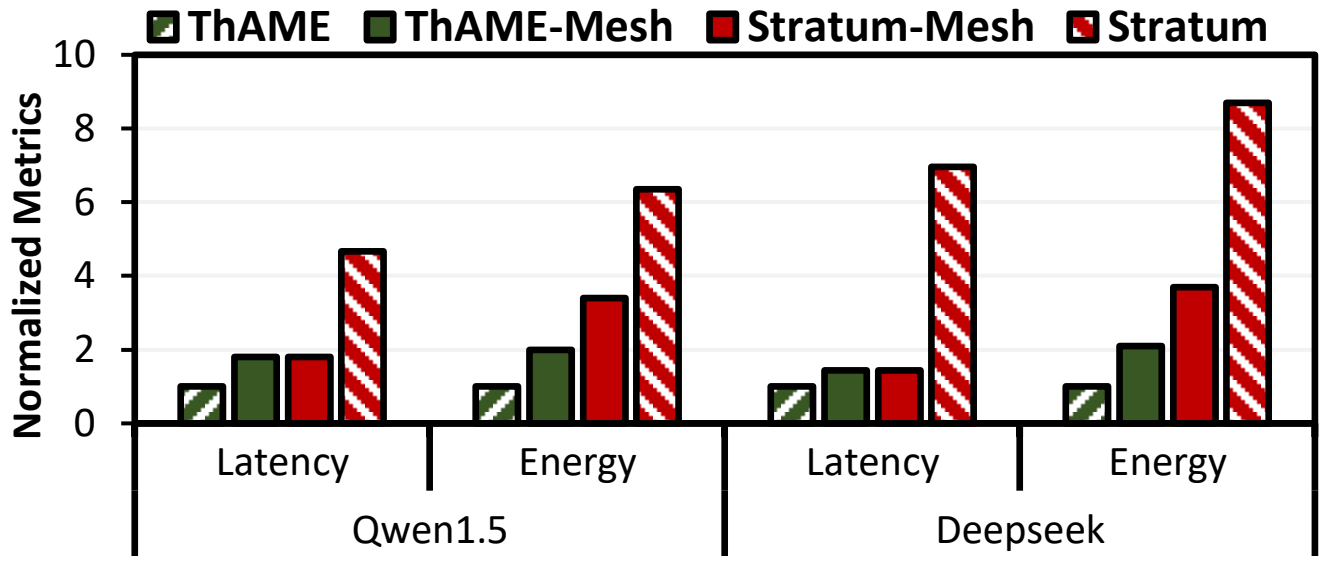

**Fig. 11:** Ablation study showing normalized expert latency and energy across different substrate and NoC configurations considering Qwen1.5 and Deepseek

ThAME with a conventional Mesh topology on the identical FeFET-NAND substrate (ThAME → ThAME-Mesh) degrades latency by up to 1.44× and energy efficiency by up to 2.1× depending on the above-mentioned models. This isolation confirms that the MOO-derived hierarchical NoC is crucial for mitigating the non-deterministic scatter-gather traffic bottlenecks. Furthermore, keeping the Mesh topology constant while substituting the FeFET-NAND substrate with DRAM-PNM (ThAME-Mesh → Stratum-Mesh) introduces an additional energy overhead up to 1.8×. The energy penalty incurred by the DRAM-PNM substrate highlights the inefficiency of continuous DRAM refresh cycles. Overall, neither the interconnect nor the memory substrate solely accounts for the total system advantage. The NoC optimization primarily drives the latency reductions by resolving network contention, whereas the FeFET-NAND substrate contributes to energy savings. The synergistic co-design of these components in ThAME yields a cumulative 4.7× and 7.0× latency improvement, alongside a 5.4× and 8.7× energy efficiency increase over Stratum for the Qwen1.5 and Deepseek models, respectively. Notably, this performance disparity widens for larger models or higher batch sizes as we show later.

### *E. Sensitivity Analysis*

The data path from the 3D FeFET-NAND array to the base die comprises two stages: the FeFET cell-level read and the Cu-Cu hybrid bonding interface. Cu-Cu hybrid bonding at comparable or finer pitch than that employed in ThAME is already in production for 3D NAND, HBM stacks, and stacked last-level caches [15], [19]. Therefore, the bonding interface is a mature data path in terms of manufacturing. Consequently, we quantify ThAME's robustness against potential device variability for the FeFET device read latency parameter. As detailed in Section IV, FeFET devices are demonstrated to achieve read latencies as low as 5 ns [23], [31]. We perform a sensitivity analysis by sweeping the FeFET device read latency from 5 ns to 100 ns, while holding the array organization and bonding interface constant. Note that ThAME conservatively assumes 10 ns read latency (see Table II). Fig. 12 presents the normalized expert latency as a function of this sweep. We observe that ThAME exhibits negligible performance degradation for read latencies up to 50 ns. This resilience is attributed to the substantial bandwidth headroom in the FeFET-NAND substrate. The array organization in ThAME provides a peak theoretical bandwidth of ~14.7 TB/s, which is over 4.4× more compared to the ~3.28 TB/s sustained bandwidth demanded by the systolic array cores (see Table V). Consequently, the device read latency must degrade by an order of magnitude to 100 ns before the expert computation latency increases by 2.2×. This analysis confirms that ThAME's performance remains robust even under significantly degraded FeFET NAND read latency.

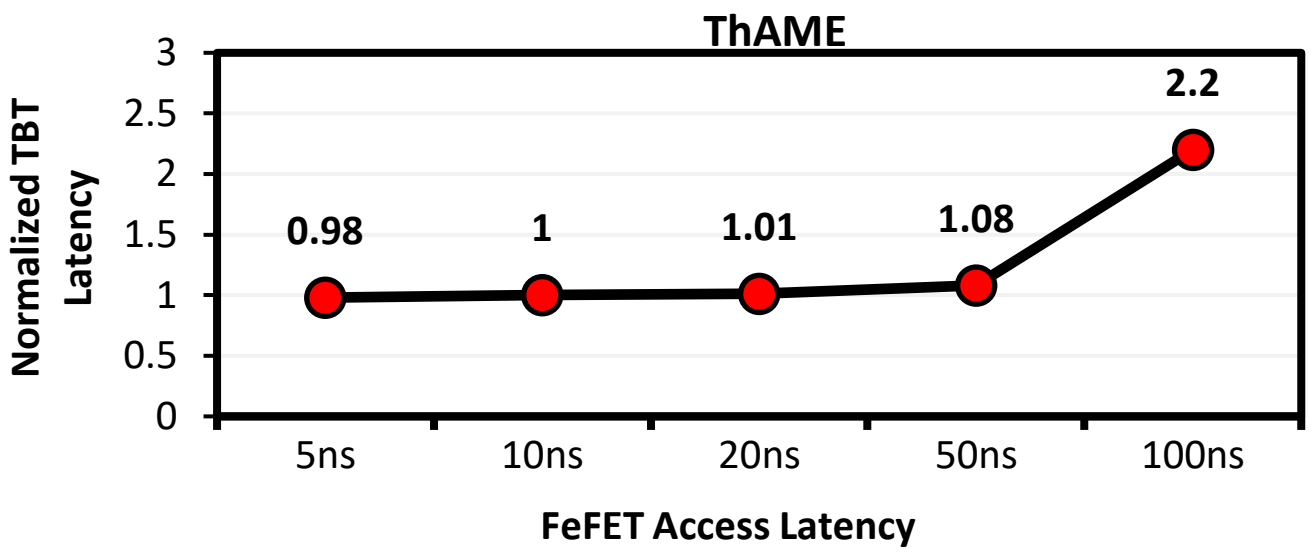

**Fig. 12:** Sensitivity of expert compute latency to FeFET device read access latency (5 ns to 100 ns).

### *F. Comparative Analysis*

The performance of LLM inference workloads is conventionally characterized by two primary metrics: Time to First Token (TTFT), which quantifies the compute-bound prefill phase, and Time Between Tokens (TBT), which governs the memory-bound decode phase. ThAME's architectural novelty explicitly targets the sparse MoE layer bottlenecks while utilizing the host for the prefill phase. Consequently, our comparative evaluation first isolates the end-to-end TBT metric and energy consumption dictated by expert processing, where ThAME's heterogeneous co-design addresses the memory and communication bottleneck. Subsequently, we report TTFT and end-to-end response time for ThAME and compare against GPU and TPU serving platforms.

We benchmark ThAME against two state-of-the-art heterogeneous accelerators: Stratum and H3D-T. Stratum mitigates the memory-bandwidth bottleneck by utilizing a monolithic DRAM-PNM substrate for expert computation. Conversely, H3D-T leverages a 3D-stacked NVM-SRAM PIM architecture. For this analysis, we consider the models detailed in Table I using a batch size ($B$) of 32 with the MMLU dataset as an example. Fig. 13(a) and (b) illustrate the normalized latency and energy, respectively, for each baseline relative to ThAME. We observe noticeably higher latency and energy for the baselines. ThAME achieves up to 10.2× and 15.7× speedup in TBT latency compared to Stratum and H3D-T, respectively, while simultaneously enhancing overall energy efficiency by 5.6× and 9.8×. H3D-T's on-chip NVM capacity is restricted to merely a few megabytes, a storage density that is inadequate for storing massive MoE experts locally. This results in frequent, high-latency accesses to off-chip storage. ThAME circumvents this bottleneck by employing the 3D FeFET-NAND chiplet which provide an aggregate on-chip capacity of 64 GB (32 GB per chiplet), accommodating the multi-billion parameter footprints of all evaluated models.

Stratum, similar to ThAME, provides significant on-chip capacity. However, its monolithic DRAM-PNM substrate suffers from three architectural constraints. First, it employs a Ring topology as the intra-chiplet interconnect, which introduces severe communication inefficiencies under non-deterministic scatter-gather traffic. Second, Stratum lacks architectural support for *Expert Parallelism*, forcing sequential processing of one expert at a time. Finally, DRAM necessitates continuous, energy-intensive refresh cycles to prevent data corruption. ThAME addresses these limitations by harnessing high-density, non-volatile FeFET-NAND memory. A fundamental limitation shared by both baselines is their NoC design to manage the dynamic scatter-gather traffic inherent to MoE token routing. ThAME mitigates this limitation by deploying a hierarchical NoC to support concurrent, multi-expert routing.

**End-to-End evaluation:** To contextualize ThAME against real-world MoE deployment platforms, we compare its end-to-end performance against NVIDIA A100 GPU and Google TPU v6e baselines. The GPU and TPU baselines are evaluated using the vLLM serving framework, with all measurements averaged over 10 iterations following a dedicated warmup phase to ensure accurate, steady-state latency and energy characterization [45]. Table VI presents the absolute TTFT and TBT in terms of latency (ms) and energy (J), alongside the end-to-end response time, formulated as TTFT $+$ $(\mathrm{n} \times \mathrm{TBT})$ for ThAME, Stratum, A100 and TPU v6e. We exclude H3D-T from this comparison since ThAME and Stratum outperform it significantly (see Fig. 13). We consider a standardized input prompt length of 128 tokens and output lengths of $n \in \{512, 1024, 2048, 4096\}$ tokens using the Qwen1.5 model with $\mathrm{B} = 32$ as a representative workload for this analysis.

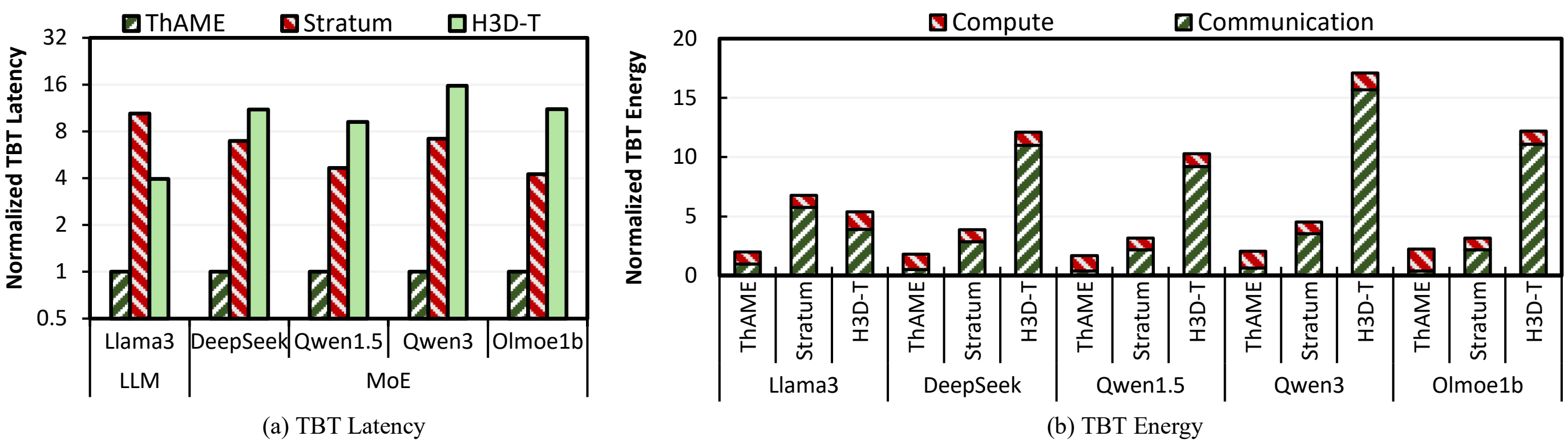

(a) TBT Latency (b) TBT Energy

**Fig. 13:** Normalized TBT (a) latency and (b) energy of the baseline architectures when compared to ThAME considering batch size of 32.

TABLE VI
ABSOLUTE TTFT, TBT, AND END-TO-END RESPONSE LATENCY AND ENERGY

| Qwen1.5 | $n = 512$ | | | $n = 1024$ | | | $n = 2048$ | | | $n = 4096$ | | |
|---|---|---|---|---|---|---|---|---|---|---|---|---|
| Architecture | TTFT (ms) | TBT (ms) | Total (ms) | TTFT (ms) | TBT (ms) | Total (ms) | TTFT (ms) | TBT (ms) | Total (ms) | TTFT (ms) | TBT (ms) | Total (ms) |
| ThAME | 89.77 | 2.17 | 1200.81 | 90.15 | 2.21 | 2353.19 | 92.42 | 2.22 | 4638.98 | 96.31 | 2.22 | 9189.43 |
| Stratum | 89.77 | 10.09 | 5256.10 | 90.15 | 10.27 | 10613.29 | 92.42 | 10.32 | 21233.92 | 96.31 | 10.32 | 42379.32 |
| A100 | 119.32 | 22.93 | 11859.4 | 120.48 | 22.93 | 23600.8 | 121.44 | 22.93 | 47082.08 | 125.42 | 22.93 | 94046.7 |
| TPU v6e | 89.77 | 18.12 | 9367.21 | 90.15 | 18.12 | 18645.03 | 92.42 | 18.12 | 37202.18 | 96.31 | 18.12 | 74315.83 |

| Qwen1.5 | $n = 512$ | | | $n = 1024$ | | | $n = 2048$ | | | $n = 4096$ | | |
|---|---|---|---|---|---|---|---|---|---|---|---|---|
| Architecture | TTFT (J) | TBT (J) | Total (J) | TTFT (J) | TBT (J) | Total (J) | TTFT (J) | TBT (J) | Total (J) | TTFT (J) | TBT (J) | Total (J) |
| ThAME | 0.673 | 0.0067 | 4.145 | 0.676 | 0.0069 | 7.74 | 0.693 | 0.0069 | 14.90 | 0.722 | 0.0069 | 29.13 |
| Stratum | 0.673 | 0.0378 | 20.047 | 0.676 | 0.0385 | 40.13 | 0.693 | 0.0387 | 79.97 | 0.722 | 0.0387 | 159.28 |
| A100 | 0.969 | 0.214 | 111.03 | 0.978 | 0.2149 | 221.25 | 0.986 | 0.2149 | 441.39 | 1.019 | 0.2149 | 881.68 |
| TPU v6e | 0.673 | 0.141 | 73.15 | 0.676 | 0.1415 | 145.66 | 0.693 | 0.1415 | 290.64 | 0.722 | 0.1415 | 580.59 |

ThAME maps the compute-bound prefill phase entirely to the host TPU chiplet, establishing an identical TTFT for ThAME, Stratum, and the standalone TPU v6e, since prefill executes on the same host device. As seen in Table VI, generative tasks impose extended decode phases that constitute the dominant fraction of overall wall-clock time. We first quantify the latency of the control-plane components detailed in section IV.D. The router (gating-network) is a $O(d_{model}.N_{experts})$ GEMM computation that completes in under 1 µs, well under 0.1% of TBT across evaluated models (see Table VI). The gating output fixes the token count, and the compute workload of each active expert. Afterwards, the scheduler divides the 32 base-die cores among the experts to balance compute latency. The scheduler solves Eq. 11 through a binary search over the latency interval $[C_{min}, C_{max}]$. The upper bound $C_{max}$ is the slowest possible expert, where all 32 tokens are routed to a single computing core (≈134,000 cycles) for Qwen1.5-MoE. The lower bound $C_{min}$ (≈17,000 cycles) is the fastest compute latency where all the tokens are equally distributed on 32 cores under uniform top-$k$ distribution. The search converges in $\lceil log_2(C_{max} - C_{min}) \rceil$ =17 iterations, varying by at most one iteration across the evaluated models. The total control-plane latency per decode step is dominated by the expert compute and synchronous gather, rendering both the gating evaluation and scheduler search practically negligible.

ThAME sustains between 2.17 and 2.22 ms TBT across all output lengths, which is 8.2× lower than TPU v6e, and 10.5× lower than A100, while consuming ≈6.7 mJ per token versus 141 mJ for TPU and 214 mJ for A100. Overall, the end-to-end speedup grows from 9.8× at 512 output tokens to 10.2× at 4096 as the one-time TTFT amortizes and the per-token TBT advantage compounds, confirming that the decode-phase gains translate to wall-clock gains for output-dominated workloads. Next, we report the compute efficiency normalized by both power and area (Tokens/s/kW/mm²) in Table VII. This metric jointly captures computational throughput, energy dissipation, and silicon area utilization. ThAME delivers 147.5 Tokens/s/W, representing a 5.6×, 20.9×, and 31.7× improvement over the Stratum, TPU v6e, and NVIDIA A100 baselines, respectively. When normalized by silicon area, ThAME achieves 184.3 Tokens/s/kW/mm², significantly outperforming the baselines. This sustained performance density confirms that the FeFET-NAND PNM substrate successfully co-locates compute with high-density storage, thereby eliminating the energy-intensive off-chip weight traffic.

TABLE VII
COMPUTE EFFICIENCY AND PERFORMANCE DENSITY COMPARISON

| | Tokens/s/W | Tokens/s/kW/mm2 |
|---|---|---|
| ThAME | 147.46544 | 184.3318 |
| Stratum | 26.427498 | 33.03437 |
| A100 | 4.6518389 | 5.16871 |
| TPU v6e | 7.0640177 | 8.830022 |

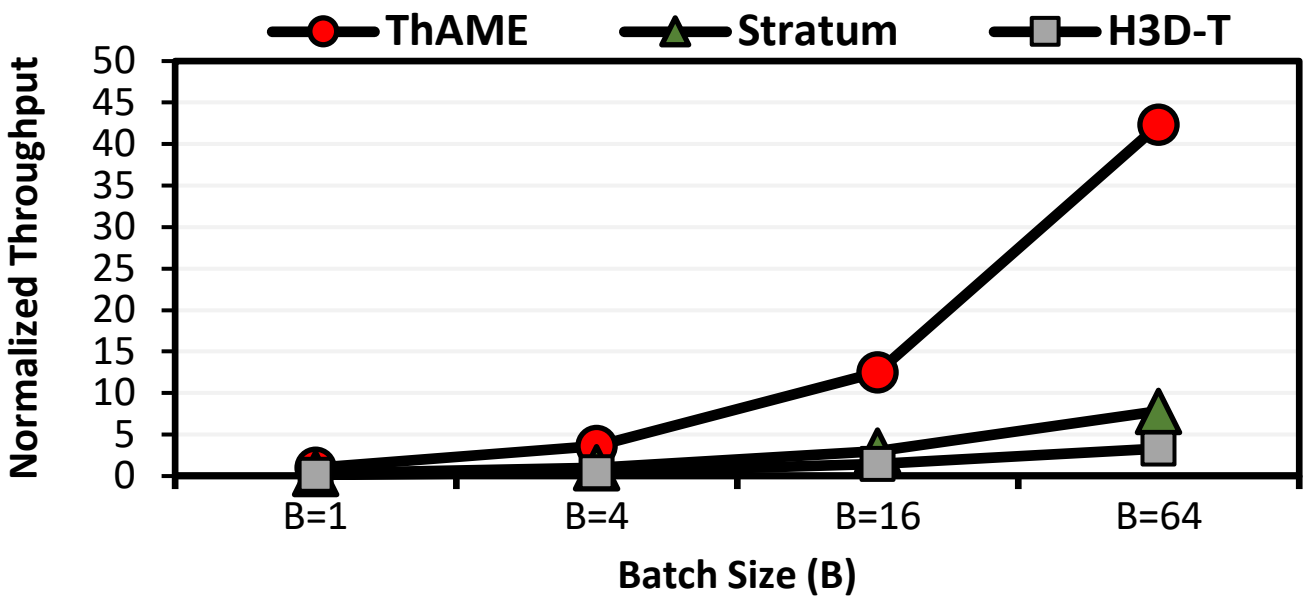

**Fig. 14:** Normalized system throughput of ThAME and baseline across varying batch sizes ($B$ in {1,4,16,64}) for the Deepseek model.

**Scalability:** Finally, we evaluate the scalability of ThAME by analyzing the normalized system throughput with varying batch size (specifically, $B \in \{1,4,16,64\}$). This scalability evaluation corresponds to request concurrency in real-world MoE LLM serving systems. As the batch size grows, the number of concurrently routed tokens exacerbates the irregularity of the scatter-gather inter-core traffic within the base die. Fig. 14 illustrates the throughput of ThAME, Stratum, and H3D-T for the Deepseek model using the MMLU dataset. It is clear that ThAME outperforms the baselines for all the batch sizes. We observe Stratum and H3D-T suffer from severe network contention and load imbalance, causing the performance difference to grow with increasing batch sizes. Specifically, ThAME delivers up to 5.92× and 12.77× higher throughput at $B = 64$ compared to Stratum and H3D-T, respectively. The results demonstrates ThAME's architectural efficiency in serving MoE workloads under heavy, concurrent batch-inference demands.

## VI. Conclusion

MoE neural architectures have emerged as a powerful paradigm for scaling LLMs, but their deployment is fundamentally bottlenecked by the extreme memory demands of static expert weights and the non-deterministic communication traffic due to input-dependent expert routing. This paper presented ThAME, a three-dimensional heterogeneous architecture designed to accelerate MoE inference. By harnessing the high storage density of 3D FeFET-NAND memory for experts, ThAME stores model parameters on-chip, effectively breaking the memory wall. Furthermore, we design a customized hierarchical NoC for the intra-chiplet traffic within FeFET-NAND. This specialized communication backbone reduces link utilization variance, mitigating the multi-hop congestion and tail-latency stalls that cripple conventional topologies such as mesh during expert processing. Experimental results demonstrate that ThAME successfully balances computation and communication under highly variable MoE traffic, achieving up to 15.7× in terms of time between tokens (TBT) latency and improving overall energy efficiency by up to 9.8× against state-of-the-art counterparts.

## REFRENCES

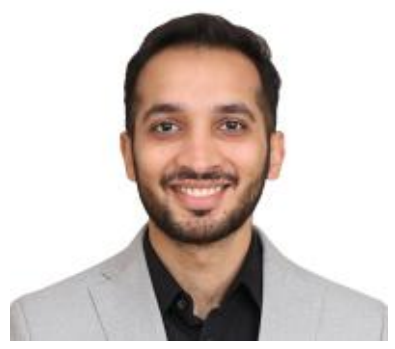

**Pratyush Dhingra** (Graduate Student Member, IEEE) is currently pursuing his Ph.D. degree in Computer Engineering at Washington State University, Pullman, WA, USA. His research interests include memory-centric architectures, AI/HPC accelerators and resource-aware ML**.**

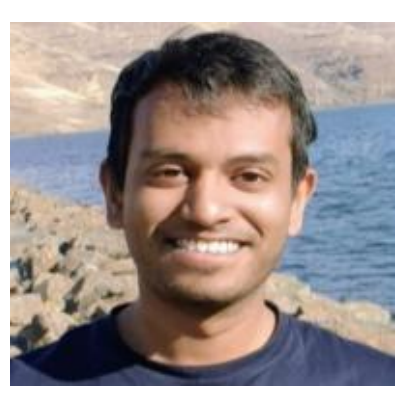

**Pramit Kumar Pal** (Graduate Student Member, IEEE) is currently pursuing his PhD from Washington State University in Pullman, WA. His research interests include AI-driven design, performance modeling, and neuromorphic computing**.**

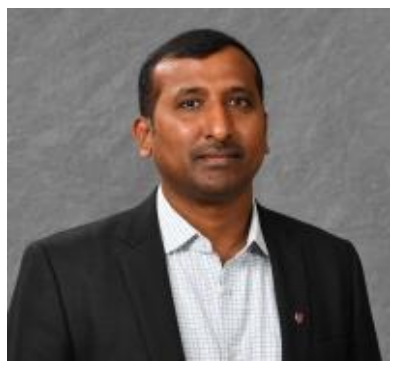

**Janardhan Rao Doppa** (Senior Member, IEEE) received the Ph.D. degree in computer science from Oregon State University, Corvallis, OR, USA, in 2014.

He is the Huie-Rogers Endowed Chair Professor of Computer Science and Berry distinguished Professor in Engineering at Washington State University, Pullman, WA, USA.

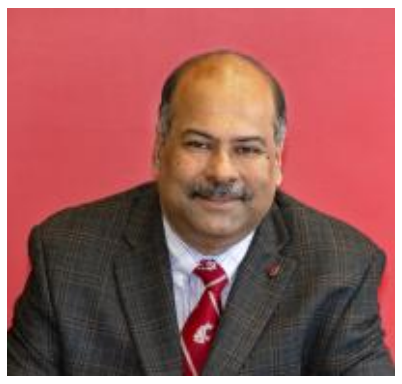

**Partha Pratim Pande** (Fellow, IEEE) received the Ph.D. degree in electrical and computer engineering from the University of British Columbia, Vancouver, BC, Canada, in 2005.

He is a Professor and Dean of Voiland College Of Engineering and Architecture at Washington State University, Pullman, WA, USA.